\documentclass{article}
\usepackage{amssymb}
\usepackage{graphicx}

\input{tcilatex}

\begin{document}

\title{Blind analysis in Physics experiments: Is this trip necessary?}
\author{R. Golub \\
Physics Dept, North Carolina State University,\\
Raleigh, NC 26695}
\maketitle

\begin{abstract}
Based on the work of Klein and Roodman [\cite{JoshK}] we present an
alternate conclusion as to the charm of blind analysis in physics
experiments.
\end{abstract}

\section{Introduction}

Blind analysis was introduced into scientific experimentation in order to
avoid the problem that in experiments with human subjects subtle clues from
the experimenter can influence the response of the subject, initiating the
dreaded placebo effect. However even double \ blinding (where both the
subject and the experimenter are unaware of \ the situation) has proved
inadequate where big pharma is involved and it has been necessary to
introduce pre-registration of clinical trials as described in \cite{JoshK}.
It has now become almost conventional to employ these techniques in physic
experiments. According to Richard Feynman \cite{feyn}

\begin{quote}
\textquotedblright It's a thing that scientists are ashamed of- this
history- because it's appparent that people did things like this: When they
got a number that was too high above Millikan's, they thought something must
be wrong - and they would look for and find a reason why something might be
wrong. When they got a number closer to Millikan's they didn't look so
hard...

The first principle is that you should not fool yourself - and you are the
easiest person to \ fool.'' \cite{feyn}

\bigskip
\end{quote}

Peter Galison \cite{Gal} has given several very interesting examples showing
how the decision to end an experiment (i.e. stop searching for systematic
errors) is influenced by previous known values of the quantity being
measured. Blind analysis would evidently help to avoid such situations.

Klein and Roodman \cite{JoshK} have given a detailed argument in support of
the need for blind analysis in nuclear and particle experiments and have
analyzed several methods of carrying this out. The abstract of their paper
summarizes the argument:

''During the past decade, blind analysis has become a widely used tool in
nuclear and particle physics experiments. A blind analysis avoids the
possibility \ of experimenters biasing their results toward their own
preconceptions by preventing them from knowing the answer until the analysis
is completer, There \ is at least citcumstantial evidence that such a bias
has affected past measurements, ans as experiments become costlier and more
difficult and hence harder to reproduce, the possibility of bias has become
a more improtant issue than in the past. We describe here the motivations
for performing a blind analysis, and give several modern exampes of
successful blind analysis strategies.''

The goal of this paper is to discuss some of the arguments in that paper

\section{The case for blind analysis and some counter-arguments}


\begin{figure}[ptb]
    \includegraphics{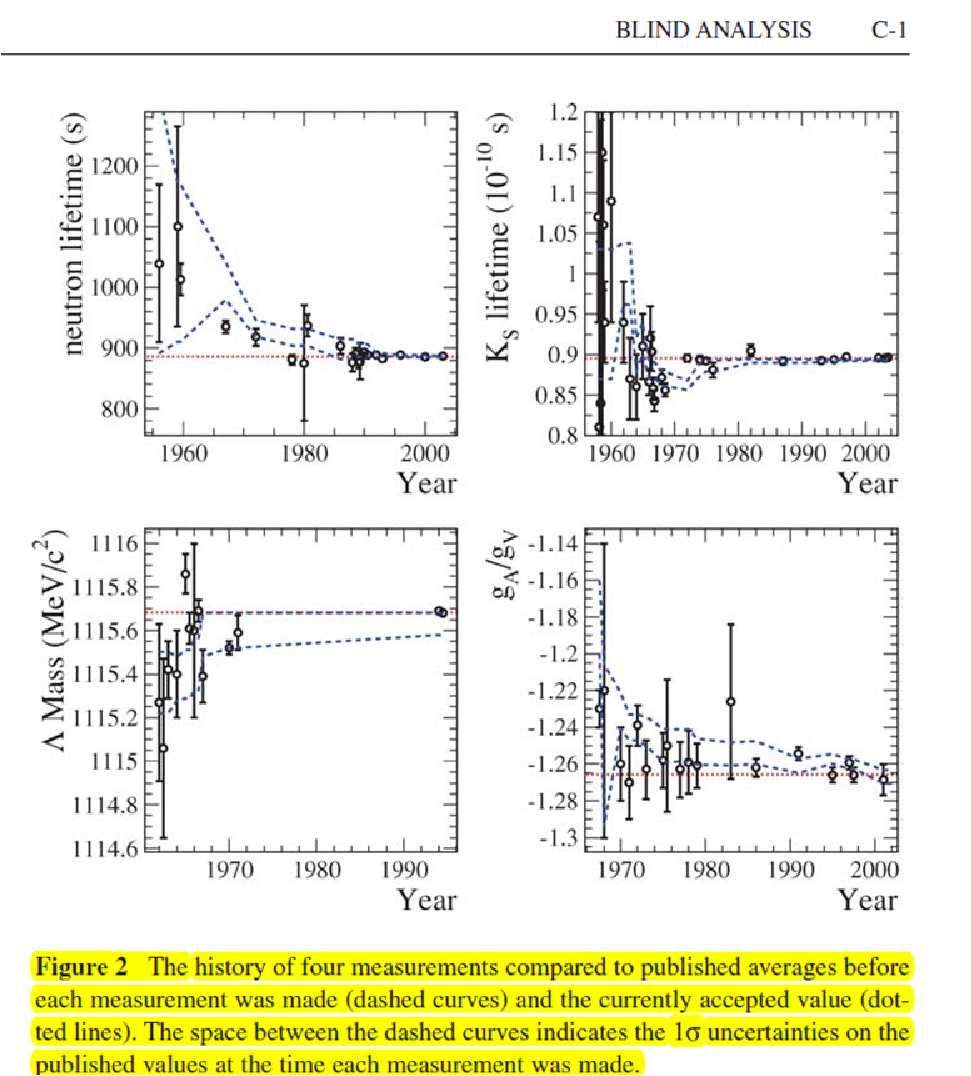}
    \caption{}
    \label{}
\end{figure}

Figure 2 of \cite{JoshK} shows historical plots of the measurements of 4
quantities. The circles are the results of individual measurements and the
dashed lines show the 1$\sigma$ limits of the published average of the
quantity at the time of measurement. The dotted horizontal lines are the
values accepted at the time of publication of \cite{JoshK}. The authors
calculate that the $\varkappa^{2}$ value for the hypothesis that the
measurements are normally distributed around the previous averages are about
1/2 those associated with the hypothesis that they are normally distributed
around the current accepted values. The dramatic shifts in value are seen to
coincide with large improvements in accuracy, indicating a radical switch in
measurement technique so that experimenter bias is not evident here. The
authors themselves conclude:

\begin{quote}
''Although we cannot say conclusively whether bias has influenced
measurements in nuclear and particle physics, the way to avoid even the
possibilities is to follow Dunnington's and Pfungst's examples and perform
measurements while staying blind to the value of our answer.''
\end{quote}

However blind analysis has real costs in that it can severely inhibit the
search for systematic errors and can preclude studies of unexpected events
that occur in 'blinded' regions of the parameter space. To introduce these
methods in an attempt to solve a problem that may be non-existent (''..avoid
even the possibility ..'') seems to the present author to be overkill. \ The
authors give an example of a case where blinding provided a serious obstacle
to the performance of an experiment:

\begin{quote}
''While looking for the decay $\pi^{+}\rightarrow e^{+}\nu$, we focused all
our attention on reducing backgrounds, since a prior experiment had set a
limit at the level of 10$^{-6}$ on the branching ratio. When we heard that
an experiment at CERN had seen a signal around 10$^{-4}$ I switched from
delayed to prompt. The signal was right there, and could have been seen on
the first day (B. Richter, private communication).''
\end{quote}

\bigskip

In addition

\begin{quote}
''We note that none of the blind techniques we describe here---and perhaps
no blind technique---can be applied to an analysis in which backgrounds are
cut or signals identified by event-by-event human inspection''.
\end{quote}

\bigskip Again quoting from \cite{JoshK}

\begin{quote}
''...what to do if the analysis breaks down...It is not necessary in the
blind analysis approach to insist that, because an analysis was done
blindly, no additional selections may be applied....The blind analysis
method does not require that data analysis stop after unblinding, nor does
it ensure that the results of the analysis are correct. There is no reason
to publish an analysis known to be wrong just because the analysis was done
blindly.

Multiple independent analyses are occasionally suggested as a way to prevent
experimenter's bias.''
\end{quote}

\bigskip

In describing a use of blinding in a measurement of the gravitational force:

\begin{quote}
''The Irvine group's measurement relied on precise knowledge of many
different detector parameters---the dimensions of the torsion balance and
test masses, the positions of the test masses, and of course the masses of
all test components. To prevent themselves from selecting data in a biased
way, or from (in their words) ``slackening of analysis effort'' when their
answer began to meet their expectations (what we have called a stopping
bias), they kept the value of their near mass known only to 1\%---the exact
mass known only to someone outside their collaboration. They used the true
value of the mass only when they had completed the analysis and were ready
to report their initial results. Subsequent improvements to the analysis
were made and later published, but they nevertheless published the
measurement made before these improvements were made.''.

So that it appears that at least in this case ''an analysis known to be
wrong'' was published ''just because the

analysis was done blindly''.
\end{quote}

\bigskip Further

\begin{quote}
''The next quandary may occur if there are more events in the signal \ box
than expected from backgrounds, \ but the events are \emph{very inconsistent
with the expected signal properties.'' }(emphasis added)

\bigskip So it appears preconceived notions cannot be completely banished.
\end{quote}

Klein and Roodman give another example of a case of blind analysis:

\begin{quote}
\bigskip''KTeV used a hidden offset directly in its $\varepsilon^{\prime
}/\varepsilon$ fit. Instead of fitting for the value of $\varepsilon^{\prime
}/\varepsilon$, the fit used%
\begin{equation}
\varepsilon^{\prime}/\varepsilon(Hidden)=\left\{ 
\begin{array}{c}
+ \\ 
-%
\end{array}
\right\} 1\times\varepsilon^{\prime}/\varepsilon+C
\end{equation}

where C was a hidden random constant, and the choice of 1 or -1 was also
hidden and random.

The +1 or -1 in the hidden value served to hide the direction $\varepsilon
^{\prime}/\varepsilon$ changed as different corrections or selections were
applied (48). In practice, KTeV had to remove the sign choice at an earlier
stage to permit a full evaluation of systematic errors. Nevertheless, the
first KTeV $\varepsilon^{\prime}/\varepsilon$ result was unblinded only one
week before the result was made public.

The addition of an unknown sign also hides the direction the result has
moved with changes to the analysis.''

So it appears that overzealous blinding (of the sign) was incompatible with
a proper consideration of systematic errors.
\end{quote}

\section{Discussion}

Based on the paper by Klein and Roodman \cite{JoshK} written in support of
the use of blind analysis in physics experiments we have attempted to show
that the method, introduced into science by medical researchers to solve a
real problem, has little place in physics research, since, as the authors
admit there is surprisingly little evidence for experimenter bias in physics
research even taking into account the cases cited by Feynman \cite{feyn} and
Galison \cite{Gal}. and the introduction of the technique has real costs in
both the ability to study systematic errors that are often unknown at the
planning stage of the experiment and come to light during the measurement
and analysis and the ability to follow up unexpected results which may only
show up after opening the box.

The experience in medical research has shown that even double blind analysis
is not sufficient to avoid experimental bias when the experimenter's are
really determined. Physicists on the other hand still have the ability to
follow the advice of my mentor, Prof Gerrold R. Zacharias who often repeated
the statement that experimental physics was really all about Character.

\end{document}